\documentclass[12pt]{article}
\usepackage{amsmath,amssymb,graphicx}

\def\mydate{December  5, 2006}
\def\ignore#1{{}}

\newcounter{sxn}

\newcounter{axn}

\date{}

\newdimen\mybaselineskip
\renewcommand{\baselinestretch}{1.25}

\renewcommand{\thefootnote}{\arabic{footnote}}

\newcommand{\beeq}{\begin{equation}}
\newcommand{\eneq}{\end{equation}}
\newcommand{\beqn}{\begin{eqnarray}}
\newcommand{\eeqn}{\end{eqnarray}}

\newcommand{\alp}{\alpha}

\newcommand{\gm}{\gamma}
\newcommand{\Gm}{\Gamma}
\newcommand{\dlt}{\delta}

\newcommand{\vep}{\varepsilon}
\newcommand{\tht}{\theta}

\newcommand{\lmd}{\lambda}

\newcommand{\sgm}{\sigma}

\newcommand{\vph}{\varphi}
\newcommand{\omg}{\omega}
\newcommand{\Omg}{\Omega}

\newcommand{\be}{\begin{equation}}
\newcommand{\ee}{\end{equation}}
\newcommand{\bea}{\begin{eqnarray}}
\newcommand{\eea}{\end{eqnarray}}
\newcommand{\eql}{&=&}

\newcommand{\defa}{&\equiv&}

\newcommand{\mtrx}[4]{\brkt{\begin{array}{cc}#1&#2\\#3&#4\end{array}}}

\newcommand{\vct}[2]{\brkt{\begin{array}{c}#1\\#2\end{array}}}

\newcommand{\tl}[1]{\tilde{#1}}
\newcommand{\bdm}[1]{{\mbox{\boldmath $#1$}}}
\newcommand{\tr}{{\rm tr}}

\newcommand{\diag}{{\rm diag}}
\newcommand{\der}{\partial}
\newcommand{\dr}{\!\!d}
\newcommand{\hc}{{\rm h.c.}}
\newcommand{\ie}{{\it i.e.}}

\newcommand{\sgn}{{\rm sgn}}

\newcommand{\brkt}[1]{\left( #1 \right)}
\newcommand{\brc}[1]{\left\{ #1 \right\}}

\newcommand{\abs}[1]{\left| #1 \right|}

\newcommand{\cD}{{\cal D}}

\newcommand{\cL}{{\cal L}}

\newcommand{\cO}{{\cal O}}

\newcommand{\zp}{z_\pi}
\newcommand{\thw}{\tht_W}
\newcommand{\thH}{\tht_{\rm H}}

\newcommand{\cph}{c_\phi}
\newcommand{\sph}{s_\phi}
\newcommand{\ubl}{U(1)_{\rm B-L}}
\newcommand{\suL}{SU(2)_{\rm L}}
\newcommand{\suR}{SU(2)_{\rm R}}
\newcommand{\uy}{U(1)_Y}
\newcommand{\mKK}{m_{\rm KK}}
\newcommand{\aL}{a_{\rm L}}
\newcommand{\aR}{a_{\rm R}}
\newcommand{\nL}{3_{\rm L}}
\newcommand{\nR}{3_{\rm R}}
\newcommand{\chL}{\pm_{\rm L}}
\newcommand{\chR}{\pm_{\rm R}}
\newcommand{\chh}{\hat{\pm}}
\newcommand{\fqR}[1]{\tl{f}^q_{R,#1}}
\newcommand{\fQR}[1]{\tl{f}^Q_{R,#1}}
\newcommand{\fqL}[1]{\tl{f}^q_{L,#1}}
\newcommand{\fQL}[1]{\tl{f}^Q_{L,#1}}

\newcommand{\NP}[1]{{\it Nucl.~Phys.}~{\bf #1}}
\newcommand{\PL}[1]{{\it Phys.~Lett.}~{\bf #1}}
\newcommand{\CMP}[1]{{\it Commun.~Math.~Phys.}~{\bf #1}}

\newcommand{\PR}[1]{{\it Phys.~Rev.}~{\bf #1}}
\newcommand{\PRL}[1]{{\it Phys.~Rev.~Lett.}~{\bf #1}}
\newcommand{\PTP}[1]{{\it Prog.~Theor.~Phys.}~{\bf #1}}

\newcommand{\AP}[1]{{\it Ann.~Phys.}~{\bf #1}}

\newcommand{\RMP}[1]{{\it Rev.~Mod.~Phys }~{\bf #1}}

\def\la{\raise.16ex\hbox{$\langle$}\lower.16ex\hbox{}  }
\def\ra{\, \raise.16ex\hbox{$\rangle$}\lower.16ex\hbox{} }
\def\go{\rightarrow}

\def\onehalf{ \hbox{${1\over 2}$} }
\def\half{ {1\over 2} }

\def\tr{{\rm tr \,}}

\def\diag{{\rm diag ~}}

\def\psibar{ \psi \kern-.65em\raise.6em\hbox{$-$} }
\def\psibarl{ \psi \kern-.65em\raise.6em\hbox{$-$} \lower.6em\hbox{} }

\begin{document}
\thispagestyle{empty}

\baselineskip=12pt

{\small \noindent \mydate    \hfill OU-HET 564/2006}


\baselineskip=35pt plus 1pt minus 1pt

\vskip 3.5cm

\begin{center}
{\Large \bf  $WWZ$, $WWH$, and $ZZH$ Couplings }\\
{\Large \bf in the Dynamical Gauge-Higgs Unification}\\
{\Large \bf in the Warped Spacetime}\\


\vspace{2.0cm}
\baselineskip=20pt plus 1pt minus 1pt

{\def\thefootnote{\fnsymbol{footnote}}
\bf 
Yutaka\ Sakamura\footnote[2]{sakamura@het.phys.sci.osaka-u.ac.jp}
and Yutaka\ Hosotani\footnote[1]{hosotani@het.phys.sci.osaka-u.ac.jp}
}\\
\vspace{.3cm}
{\small \it Department of Physics, Osaka University,
Toyonaka, Osaka 560-0043, Japan}\\
\end{center}

\vskip 2.0cm
\baselineskip=20pt plus 1pt minus 1pt

\begin{abstract}
In the dynamical gauge-Higgs unification in the Randall-Sundrum warped
spacetime, where the 4D Higgs field is unified with gauge fields and 
the electroweak symmetry is dynamically broken by the Hosotani mechanism,
the trilinear couplings for $WWZ$, $WWH$, and $ZZH$, where
$H$ stands for the Higgs field, are evaluated. 
The latter two couplings are suppressed by a factor of $\cos\thH$ 
where $\thH$ is the Yang-Mills Aharonov-Bohm phase in the extra dimension, 
while the $WWZ$ couplings remain the same as in the standard model 
to good accuracy. 
\end{abstract}



\newpage

The Higgs field in the standard model of electroweak interactions plays a vital
role in the electroweak symmetry breaking and in giving masses to
$W$, $Z$,  quarks and leptons.  The Higgs boson is expected to be discovered
at LHC in the near future.  We are entering in the era when the structure of the
electroweak symmetry breaking is disclosed.

It is not clear, however, if the Higgs sector in the standard model remains
valid at the fundamental level.  It has been argued that the Higgs boson 
mass suffers from quadratically divergent radiative corrections unless
protected by symmetry, which requires unnatural fine tuning  of parameters
in the  theory.   The leading candidate for overcoming this theoretical
unnaturalness is supersymmetry.  The minimal supersymmetric standard model
(MSSM) predicts the mass of the Higgs boson, $m_H$,  
to be less than 130 GeV.\cite{Okada}
The experimental lower bound for $m_H$ is 114 GeV.\cite{HiggsExp}

Many alternative scenarios for the Higgs sector in the electroweak interactions 
have been proposed, including the little Higgs model,\cite{littleHiggs} 
 the Higgsless model,\cite{Higgsless, Higgsless2} 
and the gauge-Higgs unification models.\cite{Fairlie1}-\cite{Carena}
Among them the dynamical gauge-Higgs 
unification predicts various properties in the Higgs field couplings and gauge 
field couplings which differ from those in the standard model and can be tested 
experimentally at LHC and future linear colliders.

In the gauge-Higgs unification scenario the  Higgs field in four dimensions 
is unified with gauge fields within the framework of higher dimensional gauge
theory.   Low energy modes of 
extra-dimensional components of gauge potentials are 4D Higgs scalar fields.
Fairlie and Manton proposed gauge-Higgs unification  in six dimensions
with  ad hoc symmetry ansatz \cite{Fairlie1, Manton1}.  
Justification for the ansatz  was attempted by making use of 
quantum dynamics,  but was afflicted with the cut-off dependence.\cite{YH3}
More attractive scheme  is obtained 
when the extra-dimensional space is non-simply connected.\cite{YH1, YH2}
There appear Yang-Mills Aharonov-Bohm phases, $\thH$, associated with
the gauge field holonomy, or the phases of Wilson line integrals along noncontractible
loops.  Although classical vacua are degenerate with respect to the values of $\thH$,
quantum dynamics of $\thH$ lifts the degeneracy
and  the non-Abelian gauge symmetry is dynamically broken.
It  is called the Hosotani  mechanism. 
Fluctuations of  $\thH$ in four dimensions correspond to the 4D Higgs  field.  
With dynamical gauge symmetry breaking,  the dynamical gauge-Higgs 
unification is achieved.

In recent years the dynamical gauge-Higgs unification has been applied to the 
electroweak interactions.  Chiral fermions are naturally accommodated in the scheme
by considering an orbifold as extra-dimensional space.\cite{Pomarol1, Antoniadis1}
  To have an $\suL$ 
doublet Higgs field, one has to start with a gauge group larger 
than $\suL \times U(1)_Y$.
As the Higgs field is a part of gauge fields, most of the couplings associated with 
the Higgs field are tightly constrained by the gauge principle.  In flat space 
$m_H$ typically turns out to be $\sim \big(g_{SU(2)}^2/4\pi  \big)^{1/2} m_W$,
which contradicts with the observation.  It is nontrivial to obtain 
quark-lepton mass matrix naturally.\cite{Csaki1}-\cite{Grzadkowski}
For instance, one needs fermions in higher dimensional representation of
the group.\cite{Csaki2}

These problems can be resolved in the dynamical gauge-Higgs unification in the
Randall-Sundrum warped spacetime.\cite{Pomarol2}-\cite{Carena}  
$m_H$ is predicted in the range 
120 GeV $\sim$ 290 GeV.  The hierarchical mass spectrum of quarks and leptons 
is naturally explained in terms of  bulk kink masses, with which couplings of 
quarks and leptons to gauge bosons and their Kaluza-Klein excited  states
are determined.  It was pointed out that the universality in the weak gauge
couplings is slightly broken, and Yukawa couplings of quarks and leptons are
substantially reduced compared with those in the standard model.\cite{HM, HNSS}

The previous model based on the gauge group $SU(3)$ is unsatisfactory
in many respects.  It gives the incorrect Weinberg angle $\theta_W$ and
the neutral current sector is unrealistic.  It is also difficult to have a realistic
fermion mass matrix.   It has been argued that dynamics at the boundaries 
(fixed points) of the orbifold such as brane kinetic terms of gauge fields can
reproduce the observed $\theta_W$.\cite{DiazCruz}

More promissing approach is to adapt a gauge group $SO(5) \times \ubl$
to start with, as advocated by Agashe, Contino and Pomarol.\cite{Agashe2}
The custodial symmetry in the 4D Higgs field sector is contained in 
$SO(5)$ and  the correct  $\theta_W$ is reproduced so that
phenomenology in the neutral currents can be reliably discussed.
In this paper we focus mainly on the gauge couplings among the gauge bosons and
the Higgs boson, to find substantial deviation from those in the standard model.
We briefly describe how interactions of fermion multiplets should be 
introduced to have realistic gauge couplings and mass matrix, but the detailed
discussions are reserved for future work.

We add that the dynamical gauge-Higgs unification is defined not only
at the tree and one-loop levels, but also beyond one loop.
It has been argued recently that the Higgs 
boson mass $m_H$, for instance, may be  finite to all order in five dimensions,
indicating that the properties of the Higgs boson can be determined 
independent of physics at the cutoff scale.\cite{Gersdorff}-\cite{YHfinite}

The model we consider is $SO(5)\times \ubl$ gauge theory in the
Randall-Sundrum (RS) geometry in five dimensions.\cite{RS1}
We use $M,N,\cdots = 0,1,2,3,4$ for the 5D curved indices, 
$A,B,\cdots=0,1,2,3,4$ for the 5D flat indices in tetrads, 
and $\mu,\nu,\cdots=0,1,2,3$ for 4D indices.
\ignore{\footnote{
As the background geometry preserves 4D Poincar\'{e} invariance, 
the curved 4D indices are not discriminated from the flat 4D indices. }}
The background metric is given by 
\be
 ds^2 = G_{MN}dx^M dx^N = e^{-2\sgm(y)}\eta_{\mu\nu}dx^\mu dx^\nu+dy^2 ~, 
 \label{metric}
\ee
where $\eta_{\mu\nu}=\diag (-1,1,1,1)$, $\sgm(y)=\sgm(y+2\pi R)$, 
and $\sgm(y)\equiv k\abs{y}$ for $\abs{y}\leq \pi R$.
The cosmological constant in the bulk 5D spacetime is given by $\Lambda= -k^2$.
$(x^\mu, -y)$ and $(x^\mu, y+ 2\pi R)$ are identified with $(x^\mu, y)$.
The spacetime is equivalent to the interval in the fifth dimension $y$ with
two boundaries at $y=0$ and $y=\pi R$, which 
we refer to  as the Planck brane and the TeV brane,  respectively. 

There are  the $SO(5)$ gauge field~$A_M$ and 
the $\ubl$ gauge field~$B_M$, the former of which  
 is decomposed as 
\be
 A_M = \sum_{I=1}^{10}A^I_M T^I 
 = \sum_{\aL=1}^3 A^{\aL}_M T^{\aL}+\sum_{\aR=1}^3 A^{\aR}_M T^{\aR}
 +\sum_{\hat{a}=1}^4 A^{\hat{a}}_M T^{\hat{a}}, 
\ee
where $T^{\aL,\aR}$ ($\aL,\aR=1,2,3$) and $T^{\hat{a}}$ ($\hat{a}=1,2,3,4$) 
are the generators of $SO(4)\sim \suL\times\suR$ and 
$SO(5)/SO(4)$,  respectively.  
As a matter field, we introduce a spinor field~$\Psi$,
belonging to the spinorial representation 
of $SO(5)$ (\ie, $\bdm{4}$ of $SO(5)$).
Extension to multi-spinor case is straightforward.  We will argue later
that multiple spinor fields are necessary to have phenomenologically
acceptable fermion content even in the one generation case.

The relevant part of the action is 
\bea
&&\hskip -1cm
 S = \int\dr^5x\;\sqrt{-G}\left [
 -\tr\brkt{\frac{1}{2}F^{(A)MN}F^{(A)}_{MN}+\frac{1}{\xi}(f^{(A)}_{\rm gf})^2
 +\cL^{(A)}_{\rm gh}}\right.    \cr
 \noalign{\kern 10pt}
 &&\hspace{0mm}
 \left.-\brkt{\frac{1}{4}F^{(B)MN}F^{(B)}_{MN}+\frac{1}{\xi}(f^{(B)}_{\rm gf})^2
 +\cL^{(B)}_{\rm gh}}
 +i\bar{\Psi}\Gm^N\cD_N\Psi-iM_\Psi\vep\bar{\Psi}{\Psi}\right ] ~, 
 \label{action1}
\eea
where $G\equiv\det(G_{MN})$, $\Gm^N\equiv e_{A}^{\;\;N}\Gm^A$. 
$\Gm^A$ is a 5D $\gm$-matrix. 
$f^{(A,B)}_{\rm gf}$ are the gauge-fixing functions, 
$\cL^{(A,B)}_{\rm gh}$ are the associated ghost Lagrangians, and 
$M_\Psi$ is a bulk mass parameter.\cite{GP}
Since the operator~$\bar{\Psi}\Psi$ is $Z_2$-odd, 
we need the periodic sign function~$\vep(y)=\sgm^\prime(y)/k$ 
satisfying $\vep(y)=\pm 1$. 
The field strengths and the covariant derivatives are defined by 
\bea
 F^{(A)}_{MN} \defa \der_M A_N-\der_N A_M-ig_A[A_M,A_N], \nonumber\\
 F^{(B)}_{MN} \defa \der_M B_N-\der_N B_M, \nonumber\\
 \cD_M\Psi \defa \brc{\der_M-\frac{1}{4}\omg_M^{\;\;AB}\Gm_{AB}
 -ig_A A_M-i\frac{g_B}{2}q_{\rm B-L} B_M}\Psi, 
\eea
where $g_A$ ($g_B$) is the 5D gauge coupling for $A_M$ ($B_M$), 
$\Gm^{AB}\equiv\frac{1}{2}[\Gm^A,\Gm^B]$, 
and $q_{\rm B-L}$ is a charge of $\ubl$. 
The spin connection 1-form $\omg^{AB}=\omg_M^{\;\;AB}dx^M$ determined from 
the metric~(\ref{metric}) is 
$ \omg^{\nu 4} = -\sgm^\prime e^{-\sgm}dx^\nu$ with all other components
vanishing. 

The boundary conditions consistent with the orbifold structure are written 
as \cite{HHHK}
\bea
&&\hskip -1cm 
 \begin{pmatrix}  A_\mu \\ A_y \end{pmatrix}  (x,-y) =
 P_0  \begin{pmatrix}  A_\mu \\ - A_y \end{pmatrix}  (x,y)  P_0^{-1} ~, \cr
 \noalign{\kern 10pt}
&&\hskip -1cm 
 \begin{pmatrix}  A_\mu \\ A_y \end{pmatrix}  (x,\pi R-y) =
 P_\pi  \begin{pmatrix}  A_\mu \\ - A_y \end{pmatrix}  (x,\pi R + y)  P_\pi^{-1} ~, \cr
 \noalign{\kern 10pt}
&&\hskip -1cm 
 \begin{pmatrix}  B_\mu \\ B_y \end{pmatrix}  (x,-y) =
 \begin{pmatrix}  B_\mu \\ - B_y \end{pmatrix}  (x,y)  ~,~
 \begin{pmatrix}  B_\mu \\ B_y \end{pmatrix}  (x, \pi R -y) =
 \begin{pmatrix}  B_\mu \\ - B_y \end{pmatrix}  (x, \pi R + y)  ~, \cr
 \noalign{\kern 10pt}
&&\hskip -.7cm 
 \Psi(x,-y) = \eta_0 P_0\gm_5\Psi(x,y) ~,~
 \Psi(x,\pi R- y) = \eta_\pi P_\pi\gm_5\Psi(x,\pi R+y) ~, 
\label{BC1}
\eea
where $\gm_5\equiv\Gm^4$ is the 4D chiral operator,
$\eta_j= \pm 1$,  $P_j \in SO(5)$ and $P_j^2 = 1$.
In the present paper, we take $P_0$ and $P_\pi$ given by 
\be
 P_0 = P_\pi = \mtrx{1_2}{}{}{-1_2} 
\label{BC2}
\ee
in the spinorial representation, or equivalently
$P_0=P_\pi = \diag (-1,-1,-1,-1,1)$ in the vectorial representation.\footnote{
For an explicit representation of the generators~$T^I$, 
see the appendix in the first reference  in Ref.~\cite{Agashe2}. }  
The boundary condition (\ref{BC1}) breaks 
 the gauge symmetry  to 
$SO(4)\times\ubl\sim \suL\times\suR\times\ubl$ at both boundaries.
(The broken generators are $T^{\hat{a}}$; $\hat{a}=1,2,3,4$.)
It is convenient to decompose $\Psi$ as 
\be
 \Psi = \vct{q}{Q},  \label{def_qQ}
\ee
where $q$ and $Q$ belong to $(\bdm{2},\bdm{1})$ and $(\bdm{1},\bdm{2})$ 
of $\suL\times\suR$, respectively. 

First notice that with (\ref{BC1}) and (\ref{BC2}) there arise zero modes
for $A_y^{\hat a}$ $(\hat a=1, \cdots, 4)$, which correspond to
the $\suL$ doublet Higgs field in the standard model;
$\Phi \propto  (A_y^{\hat 1} + i A_y^{\hat 2}, A_y^{\hat 4} - i A_y^{\hat 3})^t$.
They also give rise to Yang-Mills Aharonov-Bohm phases, $\thH$,  
in the fifith dimension.
Making use of the residual symmetry,
one can suppose that the zero mode of $A_y^{\hat 4}$
develops a nonvanishing expectation value;
\ignore{\beqn
&&\hskip -1cm
\thH =  2 g_A \int_0^{\pi R} dy \; \frac{1}{2\sqrt{2}}A_y^{\hat{4}}(y)  ~, \cr
\noalign{\kern 5pt}
&&\hskip -1cm}
\beeq
A_y^{\hat 4} =  \frac{ 2 \sqrt{2} k\,  e^{2ky}}{g_A(z_\pi^2 -1)} ~ \thH  
\label{phase1}
\eneq
where $z_\pi = e^{\pi kR}$.
Although $\thH \neq 0$ gives vanishing field strengths, it
affects physics at the quantum level.  
The effective potential for $\thH$ becomes non-trivial at the one loop level, 
whose global minimum determines the quantum vacuum.  
It is this nonvanishing $\thH$ that induces 
dynamical electroweak gauge symmetry breaking.

There are residual gauge transformations which maintain the boundary condition
(\ref{BC1}).\cite{YH2}  Among them there is a large gauge transformation given by
\beeq
\Omega^{\rm large} (y)= \exp \bigg\{ in\pi ~ \frac{e^{2ky} - 1}{z_\pi^2 - 1} ~
2\sqrt{2} \, T^{\hat 4} \bigg\} 
\label{largeGT1}
\eneq
for $0 \le y \le \pi R$ 
where $n$ is an integer.\cite{HM}  Under the transformation (\ref{largeGT1}), 
$\thH$ is  transformed to $\thH + 2\pi n$,  which implies that
all physical  quantities are periodic functions of $\thH$.  This large gauge
invariance has to be maintained at all stages.  It is vital to guarantee the
finiteness of the Higgs boson mass.\cite{YHscgt2, YHfinite}
The $\thH$-dependent part of the effective potential for $\thH$ diverges
without the large gauge invariance.\cite{HOS}

It is important to recognize that the even-odd property in (\ref{BC1}) does not 
completely fix boundary conditions of the fields. 
If there are no additional dynamics on the two branes, 
fields which are odd under parity at $y=0$ or $\pi R$  obey the Dirichlet
boundary condition (D) so that they  vanish there. 
On the other hand,  fields which are
even under parity obey the Neumann boundary conditions (N).  For gauge fields
the Neumann boundary condition is given by
$dA_\mu/dy=0$ or $d(e^{-2ky} A_y)/dy=0$.  
As a result  of additional dynamics on the branes, however,
a field with even parity, for instance,  can obey the Dirichelet boundary
condition, provided the large gauge invariance is maintained. 
We argue below that this, indeed, happens on the Planck brane.
 
Let us define new fields $A^{\prime \nR}_M$ and $A^Y_M$ by
\beqn
&&\hskip -1cm
 \vct{A^{\prime \nR}_M}{A^Y_M} =
 \mtrx{\cph}{-\sph}{\sph}{\cph} \vct{A^{\nR}_M}{B_M} ~,  \cr
\noalign{\kern 10pt}
&&\hskip -.5cm
 \cph \equiv \frac{g_A}{\sqrt{g_A^2+g_B^2}} ~~,~~
  \sph \equiv \frac{g_B}{\sqrt{g_A^2+g_B^2}} ~~.  
  \label{def_vph}
\eeqn
$A^{a_{\rm R}}_\mu$ and $B_\mu$ are even under parity, whereas
$A^{a_{\rm R}}_y$ and $B_y$ are odd.  
It is our contention that the even fields
$A^{1_{\rm R}}_\mu$, $A^{2_{\rm R}}_\mu$, and   
$A^{\prime 3_{\rm R}}_\mu$ obey the Dirichlet (D) boundary condition on the
Planck brane as a result of additional dynamics there.
The boundary conditions for gauge fields are tabulated  in Table~\ref{bd_gauge}. 
It is straightforward to confirm that the boundary conditions in Table~\ref{bd_gauge}
preserve the large gauge invariance, that is, new gauge potentials obtained by
(\ref{largeGT1}) obey the same boundary conditions as the original fields.
We note that  the Neumann   (N) boundary condition on the Planck brane 
 cannot be imposed on $A^{1_{\rm R}}_y$, $A^{2_{\rm R}}_y$, and   
$A^{\prime 3_{\rm R}}_y$, as it does not preserve the large gauge invariance.
A similar conclusion has been obtained  
in Ref.\   \cite{HNOO}.\footnote{There arises a type II defect 
on the Planck brane in the terminology of Ref.~\cite{HNOO}.}

With the condition in Table~\ref{bd_gauge}, 
 the gauge symmetry  $SO(5)\times\ubl$ in the bulk 
 is reduced to $SO(4)\times\ubl$ at the TeV brane and 
to $\suL\times\uy$ at the Planck brane.
\ignore{\footnote{
In Ref.~\cite{Agashe2} which is an earlier work 
for the Gauge-Higgs uinification in an $SO(5)\times\ubl$ model, 
the same pattern of symmetry breaking is realized 
by only the boundary conditions for the fields. 
However, such a construction leads to an inconsistent theory 
as pointed out in Ref.~\cite{HNOO}. 
In the terminology of Ref.~\cite{HNOO}, there is a type II defect 
on the Planck brane in the setup of Ref.~\cite{Agashe2}. }  }
The resultant symmetry of the theory is $SU(2)_L \times U(1)_Y$,
which is  subsequently broken to $U(1)_{\rm EM}$ by nonvanishing  
$A^{\hat{a}}_y$ ($\hat{a}=1,2,3,4$)  or $\thH$.
The weak hypercharge $Y$ is given by  $Y=T^{\nR}+q_{\rm B-L}/2$. 

One way to achieve the change of the boundary conditions of $A^{1_{\rm R}}_\mu$, 
$A^{2_{\rm R}}_\mu$, and   $A^{\prime 3_{\rm R}}_\mu$ from N to D on the Planck brane 
is to have additional fields and dynamics on the Planck brane such that 
$\suR\times\ubl$ is spontaneously broken to $\uy$ at relatively high energy 
scale~$M$,  say,  near the Planck scale~$M_{\rm Pl}$.  
Below the scale~$M$, the  mass terms \be
 \cL_{\rm mass} = 
 - \big\{  M_1^2  ( A^{1_{\rm R}}_\mu A^{1_{\rm R}\,\mu}
 +A^{2_{\rm R}}_\mu A^{2_{\rm R}\,\mu} ) 
 +M_2^2  A^{\prime \nR}_\mu A^{\prime \nR\,\mu}  \big\}   \dlt(y) ~,
 \label{L_bd1}
\ee
where $M_1, M_2 = O(M)$,  are induced on the Planck brane. 
Below the TeV scale, 
the mass terms~(\ref{L_bd1}) strongly suppress the boundary values of
($A^{1_{\rm R}}_\mu, A^{2_{\rm R}}_\mu, A^{\prime \nR}_\mu$) 
on the Planck brane, changing  the boundary conditions from N to D 
at the Planck brane. 
We note that when masses are induced by spontaneous symmetry breaking on the
Planck brane, well-controled ultra-violet behavior of gauge bosons is not 
spoiled so that the finiteness of the 4D Higgs boson mass at the one loop level, for instance,
is expected to be maintained.  It has been shown recently that the requirement of the 
tree level unitarity constrains boundary conditions satisfied by gauge 
bosons.\cite{Sakai, Higgsless, Chivukula}
With the underlying mechanism of spontaneous symmetry breaking,  the effective
boundary conditions in Table I are expected to preserve the tree level 
unitarity.

\begin{table}[t, b]
\begin{center}
\begin{tabular}{|c|c|c|c|c|}
 \hline \rule[-2mm]{0mm}{7mm} $A^{\aL}_\mu$ & $A^{1,2_{\rm R}}_\mu$ & 
 $A^{\prime \nR}_\mu$ & $A^Y_\mu$ & $A^{\hat{a}}_\mu$  \\ \hline 
 (N,N) & (D,N) & (D,N) & (N,N) & (D,D)   \\ \hline \hline
 $A^{\aL}_y$ & $A^{1,2_{\rm R}}_y$ & 
 $A^{\prime \nR}_y$ & $A^Y_y$ & $A^{\hat{a}}_y$ \\ \hline 
 (D,D) & (D,D) & (D,D) & (D,D) & (N,N) \\ \hline
 \end{tabular}
\end{center}
\caption{Boundary conditions for the gauge fields. 
$\aL,\aR=1,2,3$ and $\hat{a}=1,2,3,4$. The notation~(D,N), for example, 
denotes the Dirichlet boundary condition at $y=0$ and the Neumann boundary 
condition at $y=\pi R$. }  
\label{bd_gauge}
\end{table}

Now we  expand the bulk fields in 4D KK modes. 
It is convenient to use the conformal 
coordinate~$z\equiv e^{\sgm(y)}$ for the fifth dimension, in which
the boundaries are located at $z=1$ and  $\zp = e^{k\pi R}$. 
As in Ref.\ \cite{HNSS}, we split $A_M$ into the classical and quantum parts;
$A_M = A_M^{\rm c} + A_M^{\rm q}$.  We take $A_\mu^{\rm c}=0$ and 
$A_y^{\rm c}=(dz/dy) A_z^{\rm c}$ given by (\ref{phase1}).
Further we move to a new basis by a gauge transformation\footnote{
$\Omg(z)$ is defined   here such that $\Omg(1)=1$ whereas in 
 our previous work~\cite{HNSS} $\Omg(\zp)=1$. } 
\beqn
&&\hskip -1cm
 \tl{A}_M  =  \Omg A^{\rm q}_M \Omg^{-1}  ~~,~~ 
 \tl{B}_M = B^{\rm q}_M ~~,~~
 \vct{\tl{q}}{\tl{Q}} = z^{-2}\Omg \vct{q}{Q} ~~, \cr
\noalign{\kern 5pt}
&&\hskip -1cm
 \Omg(z) =  \exp\brc{  ig_A \int_z^{\zp}  d z^\prime ~ A^{\rm c}_z(z^\prime)} ~.
\label{basis2}
\eeqn
In the new basis the classical background of the gauge fields vanishes 
so that the linearized equations of motion reduce to the simple forms, 
while the boundary conditions become more involved. 
The 5D fields are expanded into the 4D modes as 
\bea
 \tl{A}^I_\mu(x,z) \eql \sum_n\tl{h}^I_{A,n}(z)A^{(n)}_\mu(x), \;\;\;\;\;
 \tl{A}^I_z(x,z) = \sum_n\tl{h}^I_{\vph,n}(z)\vph^{(n)}(x), \nonumber\\
 \tl{B}_\mu(x,z) \eql \sum_n\tl{h}^B_{A,n}(z)A^{(n)}_\mu(x), \;\;\;\;\;
 \tl{B}_z(x,z) = \sum_n\tl{h}^B_{\vph,n}(z)\vph^{(n)}(x), \nonumber\\
 \tl{q}_R(x,z) \eql \sum_n\fqR{n}(z)\psi_R^{(n)}(x), \;\;\;\;\;
 \tl{Q}_R(x,z) = \sum_n\fQR{n}(z)\psi_R^{(n)}(x),  \nonumber\\
 \tl{q}_L(x,z) \eql \sum_n\fqL{n}(z)\psi_L^{(n)}(x), \;\;\;\;\;
 \tl{Q}_L(x,z) = \sum_n\fQL{n}(z)\psi_L^{(n)}(x). 
 \label{md_ex}
\eea
From the action~(\ref{action1}) and the boundary conditions, 
 the mass spectrum and  analytic expressions of the mode functions are
 obtained  in terms of the Bessel functions  in the same manner as in 
  Ref.~\cite{HNSS}. 

Gauge fields are  classified  in three sectors. 
There are the charged sector (or the $W$ boson sector) 
\be
 (A^{\chL}_M,A^{\chR}_M,A^{\chh}_M) \equiv 
 \frac{1}{\sqrt{2}}(A^{1_{\rm L}}_M\pm iA^{2_{\rm L}}_M,
 A^{1_{\rm R}}_M\pm iA^{2_{\rm R}}_M,A^{\hat{1}}_M\pm iA^{\hat{2}}_M) ~, 
\ee
the neutral sector
\be
 (A^{\nL}_M,A^{\nR}_M,A^{\hat{3}}_M,B_M) ~,   \label{n_sector}
\ee
and the ``Higgs'' sector $A^{\hat{4}}_M$, which is also neutral. 
The neutral sector~(\ref{n_sector}) consists of
the $Z$ boson sector and the photon sector. 

The mass spectrum ($m=k\lambda$) in the charged sector is determined by
\beeq
F_{0,1} \Big\{ \lambda^2  z_\pi F_{0,0} F_{1,1} 
 - \frac{2}{\pi^2} \sin^2 \thH \Big\} = 0 
 \label{spectrum1}
 \eneq
where $F_{\alpha, \beta}= J_\alpha(\lambda z_\pi) Y_\beta (\lambda) -
Y_\alpha(\lambda z_\pi) J_\beta (\lambda )$.  
We note that the $\thH$-dependence appears in the form $\sin^2 \thH$
in the $SO(5) \times \ubl$ model, whereas it appeared in the form 
$\sin^2 \onehalf \thH$ in the $SU(3)$ model in \cite{HNSS}. 

The mass of the lightest mode,  the $W$ boson $W^{(0)}_\mu(x)$, 
is approximately given by 
\be
 m_W\simeq \frac{\mKK}{\pi}\sqrt{\frac{1}{k\pi R}}\abs{\sin\thH}, 
 \label{Wmass}
\ee
for $z_\pi \gg 1$ where $\mKK\equiv k\pi/(\zp-1)$ is the KK mass scale. 
Its mode functions are approximately given by 
\be
 \tl{h}^{\chL}_{A,0}(z) \simeq \frac{1+\cos\thH}{2\sqrt{\pi R}} ~, ~~
 \tl{h}^{\chR}_{A,0}(z) \simeq \frac{1-\cos\thH}{2\sqrt{\pi R}} ~, ~~
 \tl{h}^{\hat{\pm}}_{A,0}(z) \simeq \frac{\sin\thH}{\sqrt{2\pi R}}
 \brkt{\frac{z^2}{\zp^2}-1} ~. 
 \label{W_mf}
\ee

In the photon sector, there is a massless mode,
namely the photon mode~$A^{\gm(0)}_\mu(x)$, 
whose mode functions are constants. 
\bea
 \tl{h}^{\nL}_{A,0}(z) \eql \tl{h}^{\nR}_{A,0}(z) 
 = \frac{\sph}{\sqrt{(1+\sph^2)\pi R}}, \nonumber\\
 \tl{h}^{\hat{3}}_{A,0}(z) \eql 0, \;\;\;\;\;
 \tl{h}^B_{A,0}(z) = \frac{\cph}{\sqrt{(1+\sph^2)\pi R}}. 
 \label{gm_mf}
\eea
Here $\sph$ and $\cph$ are defined in (\ref{def_vph}). 
The mass spectrum in the $Z$-boson sector is determined by
\beeq
F_{0,1} \Big\{ \lambda^2  z_\pi F_{0,0} F_{1,1} 
 - \frac{2}{\pi^2}   (1+  \sph^2)  \sin^2 \thH  \Big\} = 0 ~.
 \label{spectrum2}
 \eneq
The mass $m_Z$ and the mode functions of 
the $Z$ boson $Z^{(0)}_\mu(x)$,
which is the second lightest mode in the neutral sector~(\ref{n_sector}), 
are approximately given by 
\be
 m_Z \simeq \frac{\mKK}{\pi}\sqrt{\frac{1+\sph^2}{k\pi R}}\abs{\sin\thH}, 
 \label{Zmass}
\ee
and
\bea
 \tl{h}^{\nL}_{A,1}(z) & \simeq & \frac{\cph^2+\cos\thH (1+\sph^2)}
 {2\sqrt{(1+\sph^2)\pi R}}, \;\;\;\;\;
 \tl{h}^{\nR}_{A,1}(z) \simeq \frac{\cph^2-\cos\thH (1+\sph^2)}
 {2\sqrt{(1+\sph^2)\pi R}}, \nonumber\\
 \tl{h}^{\hat{3}}_{A,1}(z) & \simeq & \sin\thH\sqrt{\frac{1+\sph^2}{2\pi R}}
 \brkt{\frac{z^2}{\zp^2}-1}, \;\;\;\;\;
 \tl{h}^B_{A,1}(z) \simeq -\frac{\sph\cph}{\sqrt{(1+\sph^2)\pi R}}. 
 \label{Z_mf}
\eea

The lightest mode~$\vph^{(0)}(x)$ in the Higgs sector  corresponds to 
the 4D Higgs field, whose  mode function  is given by 
\be
 \tl{h}^{\hat{4}}_{\vph,0}(z) = \sqrt{\frac{2}{k(\zp^2-1)}} ~ z ~. 
 \label{H_mf}
\ee
At the classical level the potential for $\vph^{(0)}$ is flat.
Quantum effects yield nontrivial, finite  corrections to the effective
potential for $\vph^{(0)}$, giving the Higgs boson  a finite mass 
$m_H=\cO(\mKK\sqrt{\alp_W kR}/(4\pi))$~\cite{HM,HNSS}.
\ignore{\footnote{
An explicit value of $m_H$ depends on the details of a model considered. } }

Some comments are in order regarding the mass spectrum determined 
by (\ref{spectrum1}) and (\ref{spectrum2}). 
First, $m_W$ and $m_Z$ are not proportional to the VEV of the ``Higgs field'', 
or $\thH$, in contrast to the ordinary Higgs mechanism. 
This is because  the Higgs mechanism, or the mechanism of mass generation, 
does not complete within each KK level 
and the lowest mode in each KK tower necessarily mixes with heavy KK modes 
when $\thH$ acquires a nonzero value. 
The similar property is seen in the $SU(3)$ model in \cite{HNSS}. 
In the $SU(3)$ model, the linear dependence of the mass spectrum 
on $\thH$ recovers 
in the flat limit, $kR\to 0$. (See Sec.~5.1 in Ref.~\cite{HNSS}.) 
In the $SO(5) \times \ubl$ model, however, the mass spectrum deviates 
from the linear dependence even in the flat limit. 
This is due to the fact that the numerical factor in front of $\sin^2\thH$ 
is a half of that in the $SU(3)$ model in \cite{HNSS}. 
It is one of the distinctive properties of the $SO(5) \times \ubl$ model. 
Secondly,  the mass spectrum of
 the modes corresponding to $F_{0,1}=0$ 
in (\ref{spectrum1}) or (\ref{spectrum2}) is independent of $\thH$. 
Their mode functions, however,  have nontrivial $\thH$-dependence.
We note that these modes do not have definite $Z_2$-parities 
when $\thH$ acquires a nonzero value, 
although the  condition $F_{0,1}=0$ is the same 
as that for the modes which have the boundary condition~(D,N) 
at $\thH=0$. 
The existence of such modes is one of the characteristics of 
the $SO(5) \times \ubl$ model. 

From (\ref{Wmass}) and (\ref{Zmass}), the Weinberg angle~$\thw$ 
determined from $m_W$ and $m_Z$ becomes
\beqn
&&\hskip -1cm
 \sin^2\thw \equiv 1-\frac{m_W^2}{m_Z^2}    \cr
\noalign{\kern 5pt}
&&\hskip .4cm
\simeq \frac{\sph^2}{1+\sph^2} = \frac{g_B^2}{g_A^2+2g_B^2}
= \frac{g_Y^2}{g_A^2 + g_Y^2} ~~. 
 \label{thW}
\eeqn
The approximate equality in the second line is valid 
to the   $\cO(0.1\%)$ accuracy for $\mKK=\cO(\mbox{TeV})$. 
In the last equality   the relation  $g_Y = g_A g_B / \sqrt{g_A^2 + g_B^2}$
has been made use of.  We note that $\sph \simeq \tan\thw$. 
The Weinberg angle $\thw$ may be determined 
from the vertices in the neutral current interactions. 
As we will see in Eqs.(\ref{gZL})-(\ref{neg_alp_gc}) below, 
$\thw$ in this definition coincides 
with that in (\ref{thW}) to good accuracy. 
Thus the rho parameter is approximately one in our model. 

\ignore{The last equality also implies that $\thw$ determined by the $W$ and $Z$ masses
nearly equals to that determined by the mixing angle of photon and $Z$,
provided that $g_Y^{4D}/g_{SU(2)}^{4D}$ in 4D is the same as $g_Y/g_A$ in 5D.
The rho parameter is approximately  one in the Randall-Sundrum warped spacetime.
implies that $\thw$ coinsides with  the  angle determined from 
the coupling constants.  The $\rho$ parameter equals to 1 within $\cO(0.1\%)$
at this level.}

The mass spectrum of a fermion multiplet (\ref{def_qQ}) is determined by 
\beeq
 \lambda^2  z_\pi F_{c-\half, c-\half} F_{c+\half , c+\half} 
 - \frac{4}{\pi^2}    
 \begin{pmatrix} \sin^2 \onehalf \thH  \cr \cos^2 \onehalf \thH \end{pmatrix}
  = 0 \quad \hbox{for ~}
  \eta_0 \eta_\pi = 
  \begin{pmatrix} +1 \cr -1 \end{pmatrix} ~.
 \label{spectrum3}
 \eneq
Here $c =  M_\Psi/k$.
The lightest mass eigenvalue~$m_f$ is approximately 
given by 
\be
 m_f \simeq k \brc{ \frac{c^2 - \frac{1}{4}}{\zp \sinh \big[ (c+\half) k\pi R \big] \, 
 \sinh \big[ (c-\half)k\pi R \big] }}^{1/2}
 \begin{pmatrix} | \sin \onehalf \thH | \cr  | \cos \onehalf \thH | \end{pmatrix} ~.
 \label{m_f}
\ee
For $c> \onehalf$ and $(\eta_0, \eta_\pi)=(1,1)$, the corresponding mode functions 
are approximately given by
\bea
 \fqL{0}(z) &\simeq& ip_{\rm H/2}
 \cos \frac{\thH}{2}  \sqrt{2k(c - \onehalf)}
~ z^{-c} ~, ~~
 \fQL{0}(z) \simeq -\abs{ \sin \frac{\thH}{2} } \sqrt{2k(c -\onehalf)} ~ z^{-c} ~, 
 \cr
\noalign{\kern 10pt}
 \fqR{0}(z) &\simeq& -i\sin\thH\frac{\sqrt{k(c+\onehalf)}}{\sqrt{2}\zp^{c+\half}}
~ z^{1-c} ~,  ~~
 \fQR{0}(z) \simeq -\frac{\sqrt{2k (c+\half) }}{\zp^{c+\half}} ~z^{c} ~, 
 \label{f_mf}
\eea
where $p_{\rm H/2}\equiv\sgn(\sin\frac{1}{2}\thH)$. 
As shown in \cite{HNSS},  the hierarchical mass spectrum for  fermions 
is obtained from (\ref{m_f}) 
by varying the dimensionless parameter $c$ in an $\cO(1)$ range. 

The $\thH$-dependence of $m_f$  differs  from that of 
$m_W$ and $m_Z$. 
Consequently the ratios~$m_f/m_W$, $m_f/m_Z$  depend on $\thH$ in the
$SO(5) \times \ubl$ model in contrast to those 
in the ordinary Higgs mechanism.\footnote{ 
In the $SU(3)$ model in \cite{HNSS}, the $\thH$-dependences of $m_f$ and $m_W$ 
are the same so that the ratio~$m_f/m_W$ becomes independent of $\thH$ 
even in the warped spacetime. }

Let us turn to the various coupling constants in the 4D effective theory. 
We first look at the 4D gauge coupling constants of fermions, 
which are obtained as overlap integrals of the mode functions. 
The result is 
\bea
 \cL^{(4)}_{\rm gc} \eql 
 \sum_n W^{(n)}_\mu\brc{\frac{g^{W(n)}_L}{\sqrt{2}}
 \bar{\psi}^{(0)}_{L2}\gm^\mu\psi^{(0)}_{L1}
 +\frac{g^{W(n)}_R}{\sqrt{2}}\bar{\psi}^{(0)}_{R2}\gm^\mu 
 \psi^{(0)}_{R1}+\hc} \nonumber\\
 &&+\sum_n Z^{(n)}_\mu \sum_{i=1}^2\brc{
 g^{Z(n)}_{Li}\bar{\psi}^{(0)}_{Li}\gm^\mu\psi^{(0)}_{Li}
 +g^{Z(n)}_{Ri}\bar{\psi}^{(0)}_{Ri}\gm^\mu\psi^{(0)}_{Ri}}
 \nonumber\\
 &&+\sum_n A^{\gm(n)}_\mu \sum_{i=1}^2
 g^{\gm(n)}_i\brc{\bar{\psi}^{(0)}_{Li}\gm^\mu\psi^{(0)}_{Li}
 +\bar{\psi}^{(0)}_{Ri}\gm^\mu\psi^{(0)}_{Ri}}
 +\cdots, 
\eea
where the index~$i=1,2$ denotes the upper or lower components 
of $\psi_{L,R}^{(0)}$, 
and the ellipsis denotes terms involving the massive KK modes. 
From the approximate expressions of the mode functions~(\ref{W_mf}),   (\ref{gm_mf}), 
(\ref{Z_mf}), and (\ref{f_mf}) for $c > \onehalf$, 
the 4D gauge couplings are found to be 
\bea
&&\hskip -1cm
 g^{W(0)}_L \simeq \frac{g_A}{\sqrt{\pi R}}\equiv g  ~,  
 \label{gWL}   \\
\noalign{\kern 10pt}
&&\hskip -1cm
 g^{Z(0)}_L \simeq  \frac{(-1)^{i-1}g_A-g_B q_{\rm B-L}\sph\cph}
 {2\sqrt{(1+\sph^2)\pi R}}
 \simeq \frac{g}{\cos\thw}\brc{\frac{(-1)^{i-1}}{2}-q_{\rm EM}\sin^2\thw}  ~,  
  \label{gZL}   \\
\noalign{\kern 5pt}
&&\hskip -1cm
 g^{\gm(0)}_i    = eq_{\rm EM} ~,  
 \label{gEM}
\eea
where $e\equiv g_A\sin\thw/\sqrt{\pi R}=g\sin\thw$ is the $U(1)_{\rm EM}$ 
gauge coupling constant and $q_{\rm EM}\equiv \brc{(-1)^{i-1}+q_{\rm B-L}}/2$ 
is the electromagnetic charge. 
The relation (\ref{thW}) has been made use of in the second equality in (\ref{gZL}). 
Note that Eqs.(\ref{gWL}), (\ref{gZL}) and (\ref{gEM}) agree with 
the counterparts in the standard model, and are consistent with 
the experimental results. 

Rigorously speaking, the couplings $g^{W(0)}_L$ and $g^{Z(0)}_L$
have small dependence on the  parameter $c$, 
which results in slight violation of the universality in weak
interactions as discussed in Ref.\ \cite{HNSS}.   It was found that there is
violation of the $\mu$-$e$ universality of $O(10^{-8})$. 

\ignore{
\bea
 g^{W(0)}_R &\simeq& \frac{g_A}{\sqrt{\pi R}}\frac{1-\cos\thw}{2} 
 = g\frac{1-\cos\thw}{2}, \label{gWR} \\
  g^{Z(0)}_R &\simeq& \frac{g_A\brc{\cph^2-\cos\thH (1+\sph^2)}
 -g_Bq_{\rm B-L}\sph\cph}{2\sqrt{(1+\sph^2)\pi R}}, 
 \label{gZR} 
 \eea
}

However, $g^{W(0)}_R$ and $g^{Z(0)}_R$ evaluated in a similar manner for the 
same multiplet $\Psi$ substantially deviate from the standard model values.
For instance, one finds $g^{W(0)}_R =  g (1 - \cos\theta_W)/2$, which is
unacceptable. 
This is because the mode functions of the right-handed fermions are localized 
near the TeV brane for $c > \onehalf$. 
Since  KK excited states are also  localized near the TeV brane, 
the mixing with KK excited states becomes strong, causing the deviation.

This implies that left-handed quarks $(u_L, d_L)$ and right-handed quarks
$(u_R, d_R)$, for instance,  cannot be in one single multiplet 
$\Psi = (q, Q)$ in (\ref{def_qQ}).    Instead one should suppose that 
$(u_L, d_L)$ is  in $q_L$ of $\Psi = (q, Q)$ with $c > \onehalf$, whereas
$(u_R, d_R)$ is in $Q_R'$ of a distinct multiplet 
$\Psi' = (q' ,   Q')$ with $c < - \onehalf$.
Indeed,  for $c < - \onehalf$, the right-handed fermions are localized 
on the Planck brane and the mixing effect  mentioned above becomes negligible. 
The couplings become 
\bea
 g^{W(0)}_R & \simeq & 0 ~, \cr
\noalign{\kern 10pt}
 g^{Z(0)}_R & \simeq & -\frac{(-1)^{i-1}g_A\sph^2+g_Bq_{\rm B-L}\sph\cph}
 {2\sqrt{(1+\sph^2)\pi R}} 
 \simeq -\frac{g}{\cos\thw}q_{\rm EM}\sin^2\thw ~, 
 \label{neg_alp_gc}
\eea
which  agree with those in the standard model.
This assignment solves another serious problem associated with fermions
localized near the TeV brane. Those fermions have too large couplings 
to the KK gauge bosons,  which may contradict with the current 
precision measurements.\cite{Chang}-\cite{Agashe}

Of course there remain additional $Q_R$ and $q_L'$ which have 
light modes.  These  modes must be made substantially heavy,   
which can  be achieved by having boundary mass terms connecting 
$Q_R$, $q_L'$ and additional boundary fields on the TeV brane.   
This issue will be discussed in more detail
in the next paper~\cite{SH2}. 

Next let us consider the trilinear couplings among the 4D gauge bosons. 
From the self-interactions of the 5D gauge fields, the following couplings 
are induced in the 4D effective theory. 
\bea
 \cL^{(4)}_{WWZ} \eql \brc{ig^{(1)}_{WWZ}
 \brkt{\der_\mu W^{(0)}_\nu-\der_\nu W^{(0)}_\mu}^\dagger W^{(0)\mu}Z^{(0)\nu}+\hc}
 \cr
\noalign{\kern 5pt}
 &&+ig^{(2)}_{WWZ}W^{(0)\dagger}_\mu W^{(0)}_\nu 
 \brkt{\der^\mu Z^{(0)\nu}-\der^\nu Z^{(0)\mu}}+\cdots. 
\eea
The couplings~$g^{(1)}_{WWZ}$ and $g^{(2)}_{WWZ}$ are expressed by 
the overlap integrals of the mode functions as
\bea
&&\hskip -1cm
 g^{(1)}_{WWZ} = g^{(2)}_{WWZ} \nonumber\\
 \noalign{\kern 5pt}
 &&\hskip .2cm
= g_A\int_1^{\zp}\frac{dz}{kz}\left[\tl{h}^{\nL}_{A,0}\brc{
 \brkt{\tl{h}^{+_{\rm L}}_{A,0}}^2+\frac{1}{2}\brkt{\tl{h}^{\hat{+}}_{A,0}}^2}
 +\tl{h}^{\nR}_{A,0}\brc{\brkt{\tl{h}^{+_{\rm R}}_{A,0}}^2
 +\frac{1}{2}\brkt{\tl{h}^{\hat{+}}_{A,0}}^2} \right. \nonumber\\
\noalign{\kern 10pt}
&&\hspace{30mm} \left.
 +\tl{h}^{\hat{3}}_{A,0}\tl{h}^{\hat{+}}_{A,0}
 \brkt{\tl{h}^{+_{\rm L}}_{A,0}+\tl{h}^{+_{\rm R}}_{A,0}}\right] ~.
\eea
Making use of  (\ref{W_mf}) and (\ref{Z_mf}), 
one finds that
\be
 g^{(1)}_{WWZ}=g^{(2)}_{WWZ}
 \simeq \frac{g_A}{\sqrt{(1+\sph^2)\pi R}}
 \simeq g\cos\thw~. 
 \label{gWWZ}
\ee
In the last equality, (\ref{thW}) and (\ref{gWL}) have been made use of. 
These couplings have the same values as those in the standard model. 
The result is consistent with the data of $e^+ e^- \go W^+ W^-$ at LEP2 
which indicates the validity of the $WWZ$ coupling in  the standard model.  
In deriving (\ref{gWWZ}), we have neglected corrections suppressed 
by a factor of $(k\pi R)^{-1} \simeq 1/35$ in conformity with the 
approximation employed in deriving  Eqs.(\ref{Wmass})-(\ref{thW}). 
The $WWZ$ couplings in the model under investigation
agree with those in the standard model within this approximation. 
Small deviation from the standard model may arise 
beyond this approximation, which needs to be evaluated numerically.
For the process $e^+ e^- \go W^+ W^-$ contributions from KK excited states
also need to be incorporated.
\ignore{
}

As the 4D Higgs field is a part of 5D gauge fields, 
the self-interactions of the 5D gauge fields also determine
the couplings of the Higgs boson~$\vph^{(0)}$ to the $W$ or $Z$ bosons
in the 4D effective theory  
\be
 \cL^{(4)} = -\lmd_{WWH} ~ \vph^{(0)}W^{(0)\mu\,\dagger}W^{(0)}_\mu 
 - \half  \lmd_{ZZH} ~ \vph^{(0)}Z^{(0)\mu}Z^{(0)}_\mu+\cdots, 
\ee
where
\bea
 \lmd_{WWH} \eql g_A k\int_1^{\zp}\frac{dz}{z}\; \tl{h}^{\hat 4}_{\vph,0}
 \brc{\tl{h}^{\hat{+}}_{A,0}\der_z\brkt{\tl{h}^{+_{\rm R}}_{A,0}
 -\tl{h}^{+_{\rm L}}_{A,0}}
 -\der_z\tl{h}^{\hat{+}}_{A,0}\brkt{\tl{h}^{+_{\rm R}}_{A,0}
 -\tl{h}^{+_{\rm L}}_{A,0}}}, \nonumber\\
 \lmd_{ZZH} \eql g_A k\int_1^{\zp}\frac{dz}{z}\; \tl{h}^{\hat 4}_{\vph,0}
 \brc{\tl{h}^{\hat{3}}_{A,0}\der_z\brkt{\tl{h}^{\nR}_{A,0}
 -\tl{h}^{\nL}_{A,0}}
 -\der_z\tl{h}^{\hat{3}}_{A,0}\brkt{\tl{h}^{\nR}_{A,0}
 -\tl{h}^{\nL}_{A,0}}}. 
\eea
With the aid of  (\ref{Wmass})-(\ref{thW}), 
these couplings are evaluated to be 
\bea
 \lmd_{WWH} & \simeq & \frac{g_A\sqrt{k}}{\pi R\zp}\sin\thH\cos\thH
\simeq gm_W\cdot p_{\rm H}\cos\thH, 
 \cr
 \noalign{\kern 10pt}
 \lmd_{ZZH} & \simeq & \frac{g_A\sqrt{k}(1+\sph^2)}{\pi R \zp}
 \sin\thH\cos\thH
 \simeq \frac{gm_Z}{\cos\thw}\cdot p_{\rm H}\cos\thH, 
 \label{lmds}
\eea
where $p_{\rm H}\equiv\sgn(\sin\thH)$. 
\ignore{We have used (\ref{Wmass}), (\ref{Zmass}), (\ref{rel_vph-thw}) and (\ref{gWL}) 
in the second equations. }
It is seen that these couplings are suppressed, compared with those in the standard 
model, by  a factor $\cos\thH$. 

\ignore{
}

So far we have neglected the $\suR$-breaking in the fermion sector. 
Since $\suR$ is  broken at the Planck brane, it is natural to have
brane-localized mass terms  with $\suR$ breaking, which, in turn, alter
mass eigenvalues and mode functions of the fermions.
%
We would like to emphasize that 
the predictions for  the gauge couplings~(\ref{gWL}), 
(\ref{gZL}), (\ref{gEM}) and  (\ref{neg_alp_gc}) remain robust 
after such an $\suR$ breaking effect is incorporated. 
The dependence of the gauge couplings on the fermion mode functions are 
exponentially suppressed as long as they are localized 
near the Planck brane.\footnote{
In this regard quantum corrections from the top quark may be important, 
and need further investigation.}
Implications of brane-localized mass terms will be  analysed in the
separate paper~\cite{SH2}. 
The presence of $SU(2)_L \times SU(2)_R$ symmetry in the bulk leads 
not only to the custodial symmetry in the 4D Higgs interactions but also
to right-handed neutrino states.  It would be interesting to implement the 
see-saw mechanism in this gauge-Higgs unification scenario.

The main result of this paper is the prediction of 
the suppression factor $\cos\thH$ for $\lmd_{WWH}$ and $\lmd_{ZZH}$.
The $WWZ$ couplings remain the standard model values.
All of these couplings can be measured at LHC and future linear colliders.
They will certainly give crucial information about the mechanism of
the symmetry breaking in the electroweak interactions.

\vskip 1cm

\leftline{\bf Acknowledgments}
The authors would like to give due thanks to  Shinya Kanemura  for
many enlightening and helpful comments.
This work was supported in part by JSPS fellowship No.\ 0509241 (Y.S.), 
and by  Scientific Grants from the Ministry of 
Education and Science, Grant No.\ 17540257,
Grant No.\ 13135215 and Grant No.\ 18204024 (Y.H.).

\vskip 1.cm

\def\jnl#1#2#3#4{{#1}{\bf #2} (#4) #3}

\def\Zphys{{\em Z.\ Phys.} }
\def\jssc{{\em J.\ Solid State Chem.\ }}
\def\jpsJ{{\em J.\ Phys.\ Soc.\ Japan }}
\def\ptps{{\em Prog.\ Theoret.\ Phys.\ Suppl.\ }}
\def\PTP{{\em Prog.\ Theoret.\ Phys.\  }}

\def\JMP{{\em J. Math.\ Phys.} }
\def\NPB{{\em Nucl.\ Phys.} B}
\def\NP{{\em Nucl.\ Phys.} }
\def\PLB{{\em Phys.\ Lett.} B}
\def\PL{{\em Phys.\ Lett.} }
\def\PRL{\em Phys.\ Rev.\ Lett. }
\def\PRB{{\em Phys.\ Rev.} B}
\def\PRD{{\em Phys.\ Rev.} D}
\def\PRe{{\em Phys.\ Rep.} }
\def\AP{{\em Ann.\ Phys.\ (N.Y.)} }
\def\RMP{{\em Rev.\ Mod.\ Phys.} }
\def\ZPC{{\em Z.\ Phys.} C}
\def\SCI{\em Science}
\def\CMP{\em Comm.\ Math.\ Phys. }
\def\MPLA{{\em Mod.\ Phys.\ Lett.} A}
\def\IJMPA{{\em Int.\ J.\ Mod.\ Phys.} A}
\def\IJMPB{{\em Int.\ J.\ Mod.\ Phys.} B}
\def\EPJC{{\em Eur.\ Phys.\ J.} C}
\def\PR{{\em Phys.\ Rev.} }
\def\JHEP{{\em JHEP} }
\def\cmp{{\em Com.\ Math.\ Phys.}}
\def\JPA{{\em J.\  Phys.} A}
\def\JPG{{\em J.\  Phys.} G}
\def\NJP{{\em New.\ J.\  Phys.} }
\def\CQG{\em Class.\ Quant.\ Grav. }
\def\ATMP{{\em Adv.\ Theoret.\ Math.\ Phys.} }
\def\ibid{{\em ibid.} }

\renewenvironment{thebibliography}[1]
         {\begin{list}{[$\,$\arabic{enumi}$\,$]}  
         {\usecounter{enumi}\setlength{\parsep}{0pt}
          \setlength{\itemsep}{0pt}  \renewcommand{\baselinestretch}{1.2}
          \settowidth
         {\labelwidth}{#1 ~ ~}\sloppy}}{\end{list}}

\def\reftitle#1{}                


\begin{thebibliography}{99}
\small
\baselineskip=14pt


\leftline{\bf References}




\bibitem{Okada}
Y.\ Okada, M.\ Yamaguchi, and T.\ Yanagida, 
\jnl{\PTP}{85}{1}{1991}.
\reftitle{Upper bound of the lightest Higgs boson mass in the minimal
supersymmetric standard model}


\bibitem{HiggsExp}
LEP Electroweak Working Group, 
http://lepewwg.web.cern.ch/lepewwg;

The Higgs Working Group at Snowmass '05,
hep-ph/0511332.
\reftitle{Toward High Precision Higgs-Boson Measurements 
at the International Linear e+ e- Collider}

\bibitem{littleHiggs}
N.\ Arkani-Hamed, A.G.\ Cohen and H.\ Georgi, 
\jnl{\PLB}{513}{232}{2001};
\reftitle{Electroweak symmetry breaking from dimensional deconstruction}

N.\ Arkani-Hamed, A.G.\ Cohen, E.\ Katz and A.E.\ Nelson, 
\jnl{\JHEP}{0207}{034}{2002};
\reftitle{The Littlest Higgs}

M.\ Schmaltz and D.\ Tucker-Smith,
\jnl{\it Ann.\ Rev.\ Nucl.\ Part.\ Sci.}{55}{229}{2005};
\reftitle{Little Higgs review}

M.\ Perelstein, 
hep-ph/0512128.
\reftitle{Little Higgs models and their phenomenology}




\bibitem{Higgsless}
C.\ Csaki, C.\ Grojean, H.\ Murayama, L.\ Pilo, and J.\ Terning, 
\jnl{\PRD}{69}{055006}{2004}.
\reftitle{Gauge theories on an interval: unitarity without a Higgs boson}

\bibitem{Higgsless2}
C.\ Csaki, C.\ Grojean, L.\ Pilo, and J.\ Terning, 
\jnl{\PRL}{92}{101802}{2004};
\reftitle{Towards a Realistic Model of Higgsless Electroweak Symmetry Breaking}

R.S.\ Chivukula, E.H.\ Simmons, H.J.\ He, M.\ Kurachi and M.\ Tanabashi,
\jnl{\PRD}{70}{075008}{2004}.
\reftitle{Structure of corrections to electroweak interactions in 
Higgsless models}



\bibitem{Fairlie1}
D.B.\ Fairlie, \jnl{\PLB}{82}{97}{1979};
\reftitle{Higgs' Fields And The Determination of The Weinberg Angle}
\jnl{\JPG}{5}{L55}{1979}.
\reftitle{Two Consistent Calculations of The Weinberg Angle}

\bibitem{Manton1}
N.\ Manton, \jnl{\NPB}{158}{141}{1979};
\reftitle{A New Six-Dimensional Approach to the Weinberg-Salam Model}

P.\ Forgacs and N.\ Manton, \jnl{\CMP}{72}{15}{1980}.
\reftitle{Space-Time Symmetries In Gauge Theories}

\bibitem{YH3}
Y.\ Hosotani,
 \jnl{\PLB}{129}{193}{1984}; 
\reftitle{Dynamical Gauge Symmetry Breaking As The Casimir Effect}
\jnl{\PRD}{29}{731}{1984}.
\reftitle{Dynamical Gauge Symmetry Breaking And Left-Right Asymmetry In Higher Dimensional Theories}



\bibitem{YH1}
Y.\ Hosotani, \jnl{\PLB}{126}{309}{1983}.
\reftitle{Dynamical Mass Generation By Compact Extra Dimensions}

\bibitem{YH2}
Y.\ Hosotani, \jnl{\AP}{190}{233}{1989}.
\reftitle{Dynamics Of Nonintegrable Phases And Gauge Symmetry Breaking}







\bibitem{Lim2}
H.\ Hatanaka, T.\ Inami and C.S.\ Lim, 
\jnl{\MPLA}{13}{2601}{1998}.
\reftitle{The Gauge Hierarchy Problem and Higher Dimensional Gauge Theories}

\bibitem{Lim1}
M.\ Kubo, C.S.\ Lim and H.\ Yamashita,
 \jnl{\MPLA}{17}{2249}{2002}.
\reftitle{The Hosotani Mechanism in Bulk Gauge Theories with an Orbifold Extra   Space $S^1/Z_2$}


\bibitem{Pomarol1}
A.\ Pomarol and M.\ Quiros, \jnl{\PLB}{438}{255}{1998};
\reftitle{The Standard Model from extra dimensions}


\bibitem{Antoniadis1}
I.\ Antoniadis, K.\ Benakli and M.\ Quiros,
\jnl{\it New. J.\ Phys.}{3}{20}{2001}.
\reftitle{Finite Higgs mass without Supersymmetry}

\bibitem{Csaki1}
C.\ Csaki, C.\ Grojean and H.\ Murayama, \jnl{\PRD}{67}{085012}{2003};
\reftitle{Standard Model Higgs From Higher Dimensional Gauge Fields}

C.A.\ Scrucca, M.\ Serone and L.\ Silverstrini, \jnl{\NPB}{669}{128}{2003}. 
\reftitle{Electroweak symmetry breaking and fermion masses from extra dimensions}

\bibitem{gaugeHiggs3}
L.J.\ Hall, Y.\ Nomura and D.\ Smith,  \jnl{\NPB}{639}{307}{2002};
\reftitle{Gauge-Higgs Unification in Higher Dimensions}

L.\ Hall, H.\ Murayama, and Y.\ Nomura, 
   \jnl{\NPB}{645}{85}{2002};
\reftitle{Wilson Lines and Symmetry Breaking on Orbifolds}

G.\ Burdman and Y.\ Nomura, \jnl{\NPB}{656}{3}{2003}; 
\reftitle{Unification of Higgs and Gauge Fields in Five Dimensions}



C.A.\ Scrucca, M.\ Serone, L.\ Silvestrini and A.\ Wulzer,
\jnl{\JHEP}{0402}{49}{2004};
\reftitle{Gauge-Higgs Unification in Orbifold Models} 

G.\ Panico and M.\ Serone, hep-ph/0502255.
\reftitle{The electroweak phase transition on orbifolds with 
gauge-Higgs unification}

\bibitem{HHKY}
N.\ Haba,  Y.\ Hosotani,  Y.\ Kawamura and T.\ Yamashita, 
\jnl{\PRD}{70}{015010}{2004};
\reftitle{Dynamical symmetry breaking in Gauge-Higgs unification on orbifold}

N.\ Haba,  K.\ Takenaga, and T.\ Yamashita, 
\jnl{\PLB}{605}{355}{2005}.
\reftitle{Partial gauge symmetry breaking via bare mass}

\bibitem{Haba}
N.\ Haba,  K.\ Takenaga, and T.\ Yamashita, 
\jnl{\PLB}{615}{247}{2005}.
\reftitle{Higgs mass in the gauge-Higgs unification}

\bibitem{HNT2}
Y.\ Hosotani, S.\ Noda and K.\ Takenaga,
\jnl{\PLB}{607}{276}{2005}.
\reftitle{Dynamical Gauge-Higgs Unification in the Electroweak Theory}

\bibitem{Csaki2}
G.\ Cacciapaglia, C.\ Csaki and S.C.\ Park,
hep-ph/0510366.
\reftitle{Fully Radiative Electroweak Symmetry Breaking}

\bibitem{Panico2}
G.\ Panico, M.\ Serone and A.\ Wulzer,
hep-ph/0510373.
\reftitle{A Model of Electroweak Symmetry Breaking from a Fifth Dimension}

\bibitem{Grzadkowski}
B.\ Grzadkowski and J.\ Wudka, 
hep-ph/0604225.
\reftitle{5-Dimensional Difficulties of Gauge-Higgs Unifications}









\bibitem{Pomarol2}
R.\ Contino, Y.\ Nomura and A.\ Pomarol, \jnl{\NPB}{671}{148}{2003}.
\reftitle{Higgs as a holographic pseudo-Goldstone boson}

\bibitem{Agashe2}
K.\ Agashe, R.\ Contino and A.\ Pomarol, 
\jnl{\NPB}{719}{165}{2005};
\reftitle{The minimal composite Higgs model}

K.\ Agashe and R.\ Contino, hep-ph/0510164. 
\reftitle{The minimal composite Higgs model and electroweak precision tests}

\bibitem{HNOO}
L.\ Hall, Y.\ Nomura, T.\ Okui and S.\ Oliver, 
\jnl{\NPB}{677}{87}{2004}.
\reftitle{Explicit supersymmetry breaking on boundaries 
of warped extra dimensions}

\bibitem{Oda1}
K.\ Oda and A.\ Weiler, \jnl{\PLB}{606}{408}{2005}.
\reftitle{Wilson Lines in Warped Space: Dynamical Symmetry Breaking and Restoration}


\bibitem{HM}
Y.\ Hosotani and M.\ Mabe, \jnl{\PLB}{615}{257}{2005}.
\reftitle{Higgs boson mass and electroweak-gravity hierarchy
from dynamical gauge-Higgs unification in the warped spacetime}


\bibitem{HNSS}
Y.\ Hosotani, S.\ Noda, Y.\ Sakamura and S.\ Shimasaki, 
\jnl{\PRD}{73}{096006}{2006}.
\reftitle{Gauge-Higgs Unification and Quark-Lepton Phenomenology 
in the Warped Spacetime}

\bibitem{Gherghetta2}
T.\ Gherghetta, hep-ph/0601213.
\reftitle{Les Houches lectures on warped models and holography}

\bibitem{Carena}
M.\ Carena, E.\ Ponton, J.\ Santiago and C.E.M.\ Wagner,
hep-ph/0607106.
\reftitle{Light Kaluza Klein States in Randall-Sundrum Models with Custodial SU(2)}


\bibitem{DiazCruz}
A.\ Aranda and J.L.\ Diaz-Cruz,  
\jnl{\PLB}{633}{591}{2006}.
\reftitle{Gauge-Higgs unification with brane kinetic terms}


\bibitem{Gersdorff}
G.v.\ Gersdorff, N.\ Irges and M.\ Quiros,
\jnl{\NPB}{635}{127}{2002}.
\reftitle{Bulk and Brane Rediative effects in Gauge Theories on Orbifolds}



\bibitem{YHscgt2}
Y.\ Hosotani, in the Proceedings of 
{\it ``Dynamical Symmetry Breaking"},  ed. M. Harada and K. Yamawaki 
(Nagoya University, 2004), p.\ 17. (hep-ph/0504272).
\reftitle{Dynamical Gauge Symmetry Breaking by Wilson Lines
in the Electroweak Theory}

\bibitem{Morris}
T.\ Morris, \jnl{\JHEP}{0501}{002}{2005}.

\bibitem{Irges}
N.\ Irges and F.\ Knechtli, 
\jnl{\NPB}{719}{121}{2005};  hep-lat/0604006.


\bibitem{Maru1}
N.\ Maru and T.\ Yamashita, hep-ph/0603237.
\reftitle{Two-loop calculation of Higgs Mass in gauge-Higgs unification:
5d Massless QED compactified on $S^1$}

\bibitem{YHfinite}
Y.\ Hosotani,  hep-ph/0607064.
\reftitle{All-order Finiteness of the Higgs Boson Mass in the Dynamical
Gauge-Higgs Unification}

\bibitem{HOS}
Y.\ Hosotani, K.\ Oda and Y.\ Sakamura,  in preparation.


\bibitem{RS1}
L.\ Randall and R.\ Sundrum,  \jnl{\PRL}{83}{3370}{1999}.


\bibitem{GP}
T.\ Gherghetta and A.\ Pomarol,
\jnl{\NPB}{586}{141}{2000}.
\reftitle{Bulk fields and supersymmetry in a slice of AdS}




\bibitem{HHHK}
N.\ Haba, M.\ Harada, Y.\ Hosotani and Y.\ Kawamura, 
\jnl{\NPB}{657}{169}{2003};   
{\it Erratum}, {\it ibid.}  B{\bf 669} (2003) {381}.
\reftitle{Dynamical Rearrangement of Gauge Symmetry on the Orbifold $S^1/Z_2$}


\bibitem{Sakai}
N.\ Sakai and N.\ Uekusa,  
hep-th/0604121.
\reftitle{Selecting gauge theories onan interval by 5D gauge transformation}

\bibitem{Chivukula}
R.S.\ Chivukula, D.A.\ Dicus, and H.-J.\ He,
\jnl{\PLB}{525}{175}{2002}.
\reftitle{Unitarity of compactified five-dimensional Yang-Mills theory}

\ignore{
C.\ Csaki, C.\ Grojean, H.\ Murayama, L.\ Pilo and J.\ Terning,
\jnl{\PRD}{69}{055006}{2004}.
\reftitle{Gauge theories on an interval: Unitarity without a Higgs boson}
}



\bibitem{Chang}
S.\ Chang, J.\ Hisano, H.\ Nakano, N.\ Okada and M.\ Yamaguchi, 
\jnl{\PRD}{62}{084025}{2000}.
\reftitle{Bulk Standard Model in the Randall-Sundrum Background}




\bibitem{Agashe3}
K.\ Agashe, A.\ Delgado, M.J.\ May and R.\ Sundrum,
\jnl{\JHEP}{0308}{050}{2003}. 
\reftitle{RS1, custodial isospin and precision tests}


\bibitem{Agashe}
S.J.\ Huber, \jnl{\NPB}{666}{269}{2003};
\reftitle{Flavor violation and warped geometry}

K.\ Agashe, G.\ Perez and A.\ Soni, \jnl{\PRD}{71}{016002}{2005}.
\reftitle{Flavor Structure of Warped Extra Dimension Models}




\bibitem{SH2}
Y.\ Sakamura and Y.\ Hosotani, in preparation.


\end{thebibliography}
\end{document}